\documentstyle[twoside,fleqn,espcrc2]{article}

\newcommand{\be}{\begin{equation}}
\newcommand{\ee}{\end{equation}}
\newcommand{\beq}{\begin{eqnarray}}
\newcommand{\eeq}{\end{eqnarray}}
\def\a{\alpha}
\def\b{\beta}
\def\g{\gamma}

\def\e{\epsilon}

\def\l{\lambda}
\def\L{\Lambda}
\def\m{\mu}
\def\n{\nu}

\def\r{\rho}

\def\w{\omega}

\def\de{\partial}
\begin{document}

\hbox{\hskip 12cm ROM2F-97/38  \hfil}
\hbox{\hskip 12cm hep-th/9711077 \hfil}
\vskip 1.8cm
\begin{center}
{\Large  \bf  Some \ Properties \ of \ Tensor \ Multiplets \\
\vskip 24pt
 in \ Six-Dimensional \ Supergravity}

\vspace{2cm}

{\large \large  Fabio Riccioni\footnote{I.N.F.N. Fellow} \ and \ Augusto Sagnotti}

\vspace{1.2cm}

{\sl Dipartimento di Fisica\\
Universit{\`a} di Roma \ ``Tor Vergata'' \\
I.N.F.N.\ - \ Sezione di Roma \ ``Tor Vergata'' \\
Via della Ricerca Scientifica, 1 \\
00133 \ Roma \ \ ITALY}
\vspace{0.4cm}
\vskip 1.2cm
{\bf Abstract}
\end{center}
\vskip 18pt
{We review some results on the complete coupling
between tensor and vector multiplets in six-dimensional $(1,0)$
supergravity.}

\vskip 2.8cm
\begin{center}
{Contribution to the Proceedings of the String Duality II Conference,
Trieste, ITALY, april 1997}
\end{center}
\vfill\eject

\title{Some Properties of Tensor Multiplets in Six-Dimensional 
Supergravity}

\author{F. Riccioni\thanks{I.N.F.N. Fellow} \ and \ A. Sagnotti\\
Dipartimento di Fisica, Universit\`a di Roma ``Tor Vergata''\\
I.N.F.N., Sezione di Roma ``Tor Vergata''\\
Via della Ricerca Scientifica, 1 \ 00133 Roma \ ITALY}

\begin{abstract}
We review some results on the complete coupling
between tensor and vector multiplets in six-dimensional $(1,0)$
supergravity.\end{abstract}

\maketitle

\section{Minimal Six-Dimensional Supergravity with Vector and Tensor
Multiplets}

In the low-energy supergravity, the Green-Schwarz mechanism \cite{gs}
results from couplings between
antisymmetric tensors and vector fields whose anomalous behavior,
for particular choices of the gauge group, cancels the reducible portion of
the anomaly generated by fermion 
loops. In ten dimensions, this restricts the possible non-abelian 
groups to two choices, 
$SO(32)$ or $E_8 \times E_8$, while the residual, factorized 
anomaly is disposed of
by the single antisymmetric tensor of the supergravity multiplet.
On the other hand, in six dimensions the irreducible part
of the gravitational anomaly cancels only provided \cite{r-d}
\be
n_H - n_V +29 n_T =273 \quad ,
\ee
where $n_V$, $n_T$ and $n_H$ are the numbers of vector, tensor and hypermultiplets,
a looser constraint compatible with a large number of non-abelian gauge groups.
Moreover, in this case one can naturally build type-I models 
with variable
numbers of tensor multiplets \cite{bs} as parameter-space orbifolds 
(orientifolds) \cite{cargese} 
of $K3$ reductions of the type-IIB string, where several
antisymmetric tensors take part in the anomaly cancellation \cite{as}. 
The two cases show another marked
difference. While the ten-dimensional Green-Schwarz coupling, 
$B \wedge (F^4- R^4)$, is a higher derivative term not visible in
the low-energy lagrangian, the gauge portion of the
corresponding six-dimensional coupling, $B \wedge (F^2 - R^2 )$, 
has sizable effects on the low-energy dynamics.

This talk is devoted to $(1,0)$ supergravity in six dimensions coupled to 
$n$ tensor multiplets and to vector multiplets. It reflects only in part the 
lecture presented by one of us (A.S.) at the ``String Duality II'' Conference,
since it expands the portion of the original discussion devoted to 
tensor multiplet couplings in six-dimensional $(1,0)$ supergravity.  
For completeness, we are also including some new results, that are fully
presented in \cite{frs}.

The starting point is the coupling of $(1,0)$ supergravity
to $n$ tensor multiplets, originally studied
by Romans \cite{romans} to lowest order in the fermi 
fields.  Nishino and Sezgin \cite{ns1} considered the coupling to
a single tensor multiplet and to vector and hypermultiplets to all orders in the
fermi fields, and have recently included some higher-order fermionic
couplings in the theory with tensor, vector and hypermultiplets \cite{ns2}. 
Here we complete the field equations and relate them to Wess-Zumino consistency
conditions, in the spirit of \cite{fms}, for the case of an arbitrary number
of tensor and vector multiplets. The $n$ scalars in the tensor 
multiplets parameterize the coset space $SO(1,n)/SO(n)$, and are described
by the $SO(1,n)$ matrix \cite{romans}
\be
V =\pmatrix{v_r \cr x^m_r}\quad .
\ee
All spinors are symplectic Majorana-Weyl. In particular, the tensorini $\chi^m$ 
are right-handed, while the gravitino $\Psi_\m$ and the gaugini $\l$
are left-handed. The tensor fields $B^r_{\m\n}$ are valued in the fundamental 
representation of $SO(1,n)$, and their field strengths include
Chern-Simons 3-forms of the vector fields according to
\be
H^r  = d B^r -c^{rz} \w_z \quad ,
\ee
where the $c^{rz}$ are constants that determine the gauge part of the
residual anomaly polynomial
\be
{\cal A} = \sum_{rs} \eta_{rs} c^{rx} c^{sy} tr_x ( F \wedge F ) \ 
tr_y ( F \wedge F )
\ee
and $z$ runs over the various factors of the gauge group \cite{as}. 
Gauge invariance of $H^r$ then requires that 
\be
\delta B^r = c^{rz} tr_z (\L dA) \quad .
\ee

To lowest order in the fermi fields, the fermionic equations are \cite{as}
\beq
& & \g^{\m\n\r} D_\n \Psi_\r + v_r H^{r \m\n\r} \g_\n \Psi_\r\nonumber\\
&&-\frac{i}{2} x^m_r H^{ r \m\n\r} \g_{\n\r} \chi^m  
\nonumber\\
& &+ \frac{i}{2} x^m_r \de_\n v^r \g^\n \g^\m \chi^m \nonumber\\
&&-\frac{1}{\sqrt{2}}
v_r c^{rz} tr_z (F_{\sigma \tau} \g^{\sigma \tau} \g^\m \l )=0 \quad ,
\eeq
\beq
& & \g^\m D_\m \chi^m -\frac{1}{12} v_r H^{r \m\n\r} 
\g_{\m\n\r} \chi^m\nonumber\\
&&-\frac{i}{2} x^m_r H^{r \m\n\r} \g_{\m\n} \Psi_\r \nonumber\\
& & - \frac{i}{2} x^m_r \de_\n v^r \g^\m \g^\n \Psi_\m \nonumber\\
&&-\frac{i}{\sqrt{2}}
x^m_r c^{rz} tr_z (F_{\m\n}\g^{\m\n} \l ) =0 
\eeq
and
\beq
& & v_r c^{rz}\g^\m D_\m \l +\frac{1}{2} (\de_\m v_r ) c^{rz} \g^\m \l\nonumber\\
&&+\frac{1}{2\sqrt{2}} v_r c^{rz} F_{\a\b} \g^\m \g^{\a\b} \Psi_\m \nonumber\\
& &  +\frac{i}{2\sqrt{2}}x^m_r c^{rz} F_{\m\n} \g^{\m\n} 
\chi^m \nonumber\\
&&-\frac{1}{12} c^{rz} H_{r \m\n\r} \g^{\m\n\r} \l =0 \quad ,
\eeq
while the bosonic equations are
\beq
& & R_{\m\n} -\frac{1}{2} g_{\m\n} R + \de_\m v^r \de_\n v_r \nonumber\\
&&-\frac{1}{2} g_{\m\n} \de_\a v^r \de^\a v_r -
G_{rs} H^r_{\m\a\b} H^s{}_\n{}^{\a\b} \nonumber\\
& &+ 4 v_r c^{rz}
tr_z (F_{\a\m} F^\a{}_\n \nonumber\\
& & -\frac{1}{4} g_{\m\n} F_{\a\b}F^{\a\b})=0\quad ,
\eeq
\beq
&&x^m_r \nabla_\m (\de^\m v^r ) +\frac{2}{3} x^m_r v_s H^r_{\a\b\g} 
H^{s \a\b\g} \nonumber\\
&&-x^m_r c^{rz} tr_z (F_{\a\b} F^{\a\b})=0 \label{vectorcov}
\eeq
and
\be
\nabla_\m (v_r c^{rz} F^{\m\n} ) -c^{rz} G_{rs} H^{s \n\r \sigma} F_{\r\sigma}=0
\quad .\label{vectoreq}
\ee
Moreover, the tensor fields satisfy (anti)self-duality conditions, conveniently
summarized as \cite{fms}
\be
G_{rs} H^{s \m\n\r} =\frac{1}{6e} \e^{\m\n\r\a\b\g} H_{r \a\b\g}\quad ,
\label{selfdual}
\ee
where $G_{rs} =v_r v_s + x^m_r x^m_s$. Under the supersymmetry 
transformations
\beq
& & \delta e_\m{}^a =-i(\bar{\e} \g^a \Psi_\m ) \quad,\nonumber\\
& & \delta B^r_{\m\n} =i v^r \bar{\Psi}_{[\m} \g_{\n]} \e + \frac{1}{2}
x^{mr} \bar{\chi}^m \g_{\m\n} \e \nonumber\\
&&-2c^{rz} tr_z (A_{[\m}\delta A_{\n]}) \quad,
\nonumber\\
& & \delta v_r = x^m_r \bar{\chi}^m \e \quad,\nonumber\\
& & \delta A_\m = -\frac{i}{\sqrt{2}} (\bar{\e} \g_\m \l ) \quad ,\nonumber\\
& & \delta \Psi_\m =D_\m \e +\frac{1}{4} v_r H^r_{\m\n\r}
\g^{\n\r}\e \quad ,\nonumber\\
& & \delta \chi^m = \frac{i}{2} x^m_r \de_\a v^r  \g^\a \e +
\frac{i}{12} x^m_r H^r_{\a\b\g} \g^{\a\b\g}\e \quad ,\nonumber\\
& & \delta \l =-\frac{1}{2\sqrt{2}} F_{\m\n} \g^{\m\n} \e \quad,
\eeq
the fermionic field equations turn into the bosonic ones, and 
this proves supersymmetry to lowest order.

This theory has a gauge anomaly, as one can see from the divergence of the  
(non-integrable) vector equation \cite{fms}
\beq
&&\nabla_\m J^\m = \nonumber\\
&&-\frac{1}{2e} \e^{\m\n\a\b\g\delta} c^{rz}c_r^{z^\prime}
F_{\m\n} tr_{z^\prime}(F_{\a\b} F_{\g\delta} ) \quad , \label{covanomaly}
\eeq
that reproduces the residual covariant anomaly. An integrable equation,
\beq
& & \nabla_\m (v_r c^{rz} F^{\m\n} ) -  c^{rz} G_{rs} H^{s \n\r\sigma}
F_{\r\sigma}  \nonumber\\
& & - \frac{1}{8e} \e^{\n\r\a\b\g\delta} c_r^z A_\r c^{rz^\prime} 
tr_{z^\prime} (F_{\a\b}F_{\g\delta}) \nonumber\\
& & -\frac{1}{12e}\e^{\n\r\a\b\g\delta} c_r^z F_{\r\a} c^{rz^\prime}
\w^{z^\prime}_{\b\g\delta} = 0 \quad, \label{newvectoreq}
\eeq
defines a gauge current that
generates the residual consistent anomaly. However, the resulting 
theory is no longer supersymmetric, but has a corresponding 
supersymmetry anomaly \cite{fms}
\beq
{\cal{A}}_\e = - c^{rz} c_r^{z^\prime } [
tr_z (\delta_\e A A ) tr_{z^\prime} (F^2 ) \nonumber
\eeq
\be
-2 tr_z (\delta_\e AF )
\w^{z^\prime}_3 ]\quad ,\label{susyan}
\ee
determined by the Wess-Zumino 
consistency conditions \cite{wz}. Moreover, one 
can prove that the divergence of 
the gravitino equation yields the supersymmetry anomaly. 

As usual, the commutators of supersymmetry transformations close only on-shell
on fermi fields. In order to complete the construction, {\it all}
first order field equations and the supersymmetry transformations of the
fermi fields must be made supercovariant. It should be appreciated that the 
former include the (anti)self-duality  conditions (\ref{selfdual}) on tensor 
fields \cite{schwarz}. The complete theory can then be 
constructed imposing closure of the algebra on the fermi fields.
This requires a long and tedious calculation, with the end result that the 
algebra closes on-shell in the expected fashion on all fermi fields, with the 
notable exception of the gaugini. Indeed, in this case the algebra generates
all the local symmetries and the field equation, together with an additional 
term
\beq
& & [\delta_1 , \delta_2 ]_{extra} \l =\nonumber\\
&& \frac{c \cdot c^\prime}{v \cdot c}
tr_{z^\prime} [-\frac{1}{4}(\bar{\e}_1 \g_\a \l^\prime )(\bar{\e}_2
\g_\b \l^\prime ) \g^{\a\b} \l \nonumber\\
& & +\frac{1}{4}(\bar{\l} \g_\a \l^\prime )(\bar{\e}_1 \g^\a \l^\prime )
\e_2 -(1 \leftrightarrow 2) \nonumber\\
& & +\frac{1}{16} (\bar{\e}_1 \g^\a \e_2 )(\bar{\l}^\prime \g_{\a\b\g}
\l^\prime ) \g^{\b\g} \l ] \quad .\label{alfa=0}
\eeq
We will return in the next Section to the crucial role of this charge.

From the complete fermionic equations, one can deduce a lagrangian
that generates them, and finally derive from it the complete bosonic
equations. The lagrangian is
\beq
&&e^{-1}{\cal{L}}=-\frac{1}{4}R -\frac{1}{2} v_r c^{rz} tr_z (F_{\a\b} F^{\a\b})
\nonumber\\
&&+\frac{1}{12}G_{rs} H^{r \m\n\r} 
H^s_{\m\n\r}
-\frac{1}{4} \de_\m v^r \de^\m v_r\nonumber\\ 
&&-\frac{1}{8e} \e^{\m\n\a\b\g\delta} c_r^z B^r_{\m\n} tr_z (F_{\a\b}
F_{\g\delta}) \nonumber\\
&&-\frac{i}{2}\bar{\Psi}_\m \g^{\m\n\r} D_\n \Psi_\r 
-\frac{i}{2}v_r H^{r \m\n\r}(\bar{\Psi}_\m \g_\n \Psi_\r)
\nonumber \\
& & +\frac{i}{2} \bar{\chi}^m \g^\m D_\m
\chi^m -\frac{i}{24}v_r H^r_{\m\n\r} (\bar{\chi}^m \g^{\m\n\r} \chi^m )\nonumber
\\ &&+\frac{1}{2}x^m_r \de_\n v^r (\bar{\Psi}_\m \g^\n
\g^\m \chi^m) \nonumber\\
& & -\frac{1}{2} x^m_r H^{r \m\n\r} (\bar{\Psi}_\m \g_{\n\r} \chi^m )\nonumber\\
& & +\frac{i}{\sqrt{2}} \bar{\Psi}_\m \g^{\sigma \delta}\g^\m v_r
c^{rz} tr_z [F_{\sigma \delta} \l ] \nonumber\\
& & +\frac{1}{\sqrt{2}}x^m_r c^{rz} tr_z (\bar{\chi}^m \g^{\m\n}
\l F_{\m\n} )\nonumber\\
& & +iv_r c^{rz} tr_z (\bar{\l} \g^\m D_\m \l ) \nonumber\\
&& +\frac{i}{12}
x^m_r x^m_s H^r_{\m\n\r} c^{sz} tr_z (\bar{\l}\g^{\m\n\r} \l ) \quad ,
\label{lag}
\eeq
up to terms quartic in the fermions that we have omitted for simplicity. 
Actually, this is not quite the proper lagrangian
of the theory, that would involve some peculiar couplings 
of the antisymmetric tensors,
following the original proposal of Pasti, Sorokin 
and Tonin \cite{pst}, necessary to enforce their 
(anti)self-duality conditions.
However, using these only 
after varying, eq. (\ref{lag}) yields the proper equations of all fields
aside from the antisymmetric tensors. Moreover, varying with respect 
to the antisymmetric tensors yields their second-order equations,
namely the divergence of the (anti)self-duality conditions of eq. (\ref{selfdual}).
The vector equation is not covariant already 
to lowest order, but all terms of higher order in
the fermi fields do not affect the gauge anomaly.
 
It is now straightforward to study the effect of supersymmetry transformations,
and one obtains
\be
\delta B (eq. \ B ) +\delta F (eq. \ F)={\cal A}_\e\quad ,\label{deltaL}
\ee
where $F$ and $B$ denote collectively the fermi and bose fields, aside from 
the antisymmetric tensors. Moreover, in eq. (\ref{deltaL}) one has 
to impose the complete (anti)self-duality conditions 
on the 3-form field strengths.
The complete supersymmetry anomaly ${\cal A}_\e$ that solves the Wess-Zumino
consistency conditions comprises eq. (\ref{susyan}), 
together with the additional terms
\beq
&& \Delta {\cal A}_\e =\nonumber\\
&& e c\cdot c^\prime tr \lbrace \frac{i}{2} \delta_\e A_\m F_{\n\r} 
(\bar{\l}^\prime \g^{\m\n\r} \l^\prime )\nonumber\\
&&+\frac{i}{2}  \delta_\e A_\m (\bar{\l} \g^{\m\n\r} \l^\prime )
F^\prime_{\n\r} \nonumber\\
& & + i \delta_e A_\m (\bar{\l}\g_\n \l^\prime ) F^{\prime \m\n}\nonumber\\
&&+\frac{1}{32} \delta_\e e_\m{}^a (\bar{\l} \g^{\m\a\b} \l )(\bar{\l}^\prime
\g_{a \a\b} \l^\prime ) \nonumber\\
& & -\frac{1}{2\sqrt{2}} \delta_\e A_\a (\bar{\l} \g^\a \g^\b \g^\g 
\l^\prime )(\bar{\l}^\prime \g_\b \Psi_\g ) \nonumber\\
&&+\frac{x^m_s c^{s z^\prime}}{v_t c^{t z^\prime}} [-\frac{3i}{2\sqrt{2}}
\delta_\e A_\a (\bar{\l} \g^\a \l^\prime )(\bar{\l}^\prime \chi^m )
\nonumber\\
& &- \frac{i}{4 \sqrt{2}}\delta_\e A_\a (\bar{\l} \g^{\a\b\g} \l^\prime
)(\bar{\l}^\prime \g_{\b\g}\chi^m )\nonumber\\
&&-\frac{i}{2\sqrt{2}}  \delta_\e A_\a 
(\bar{\l} \g_\b \l^\prime )(\bar{\l}^\prime \g^{\a\b} \chi^m )]\rbrace\quad .
\eeq

In general, the anomaly is defined up to the variation of a local functional. 
In our case, this reflects the freedom of adding to the lagrangian 
the term
\be
{\cal{L}}_{\l^4} = e \frac{\a}{2} c_r^z c^{r z^\prime} tr_{z, z^\prime}
[(\bar{\l} \g^\a \l^\prime )(\bar{\l}\g_\a \l^\prime ) ] \quad,\label{lambda4}
\ee
that modifies the supersymmetry anomaly according to
\be 
\Delta^\prime {\cal A}_\e = \delta_\e {\cal L}_{\l^4}\quad .\label{extraanomaly}
\ee
This term affects the field equation of the 
gaugino, and thus the commutator of eq. (\ref{alfa=0}) has to be reconsidered
in order to accommodate this modification.  As a result, after using the
modified field equation, one is left with the residual term
\beq
& & [\delta_1 , \delta_2 ]_{extra} \l =\nonumber\\
&& \frac{c \cdot c^\prime}{v \cdot c}
tr_{z^\prime} [-\frac{1}{4}(\bar{\e}_1 \g_\a \l^\prime )(\bar{\e}_2
\g_\b \l^\prime ) \g^{\a\b} \l \nonumber\\
& & -\frac{\a}{2} (\bar{\l} \g_\a \l^\prime )(\bar{\e}_1 \g_\b \l^\prime )
\g^{\a\b} \e_2 \nonumber\\
&&+\frac{\a}{16}(\bar{\l}\g_{\a\b\g}\l^\prime )(\bar{\e}_1 \g^\g 
\l^\prime ) \g^{\a\b} \e_2 \nonumber\\
& & +\frac{\a}{16} (\bar{\l} \g_\g \l^\prime )(\bar{\e}_1 \g^{\a\b\g} \l^\prime
) \g_{\a\b} \e_2 \nonumber\\
&&+\frac{1-\a}{4} (\bar{\l} \g_\a \l^\prime )
(\bar{\e}_1 \g^\a \l^\prime ) \e_2 -(1 \leftrightarrow 2) \nonumber\\
& &+\frac{1-\a}{16} (\bar{\e}_1 \g^\a \e_2 )(\bar{\l}^\prime
\g_{\a\b\g} \l^\prime  )\g^{\b\g}\l ]\quad .\label{centralcharge}
\eeq
As we will see in the next section, the presence of this 2-cocycle is 
strictly correlated to the presence of the anomaly, and indeed one
can not eliminate it, as befits a theory that 
has a built-in anomaly  for every value of $\a$.
Finally, one may show that the divergence of the complete gravitino equation 
is proportional to the complete supersymmetry anomaly.

\section{Comments on Gauge and Supersymmetry Anomalies}

If a supersymmetric Yang-Mills theory formulated in the Wess-Zumino gauge has 
a gauge anomaly, it must also have a supersymmetry anomaly, on account of
the Wess-Zumino consistency conditions
\beq
& & \delta_\e {\cal{A}}_\L = \delta_\L {\cal{A}}_\e \quad ,\nonumber \\
& & \delta_{\e_1} {\cal{A}}_{\e_2} -\delta_{\e_2} {\cal{A}}_{\e_1}=
{\cal{A}}_{\tilde{\L}} \quad ,\label{wesszumino}
\eeq
where $\tilde{\L}$ denotes the gauge parameter
generated by the commutator of two supersymmetry transformations.
The simplest example is provided by the globally supersymmetric Yang-Mills 
theory in four dimensions. From the four-dimensional gauge anomaly
\be
{\cal{A}}^4_\L =tr [\L (dA)^2 +\frac{i}{2} g d\L A^3 ] \quad,
\ee
and from eqs. (\ref{wesszumino}), one can determine the form of
the corresponding supersymmetry anomaly \cite{wzsusy}. 
In particular, from the first of eqs. (\ref{wesszumino}) one obtains 
\be
{\cal{A}}^4_\e =tr [\delta_\e A A (dA) +\delta_\e A(dA) A-\frac{3ig}{2}\delta_\e
A A^3 ]\ .
\ee
This satisfies the second of eqs. (\ref{wesszumino}) if augmented
 by the gauge-invariant term 
\be
\Delta {\cal{A}}_\e^4 = -\frac{i}{2}tr [\delta_\e A \bar{\l} \g^{(3)} \l
+ \bar{\l} \delta_\e A \g^{(3)} \l ] \quad ,
\ee
where $\g^{(3)}=\g_{\m\n\r}dx^\m \wedge dx^\n 
\wedge dx^\r$, $\l$ is a right-handed Weyl spinor and
\beq
\delta A_\m = \frac{i}{\sqrt{2}}(\bar{\e} \g_\m \l - \bar{\l} \g_\m \e )\quad ,
\nonumber
\eeq
\be
\delta \l =\frac{1}{2\sqrt{2}} F_{\m\n} \g^{\m\n} \e \quad .
\ee

Although the algebra closes only on the field equation
of $\l$, in four dimensions it is not possible to generate
from eqs. (\ref{wesszumino}) a term proportional to $\g^\m D_\m \l$,
and thus the Wess-Zumino consistency conditions close also off-shell. This
was originally observed in \cite{wzsusy}. 

On the other hand, in six dimensions, starting from the residual gauge anomaly 
\be
{\cal{A}}^6_{\L} = - c^{rz} c_r^{z^\prime} tr_z (\L dA ) tr_{z^\prime}
(F^2 ) \quad ,
\ee
the first of eqs. (\ref{wesszumino}) implies a 
corresponding supersymmetry anomaly
\beq
{\cal{A}}^6_\e = - c^{rz} c_r^{z^\prime } [
tr_z (\delta_\e A A ) tr_{z^\prime} (F^2 ) \nonumber
\eeq
\be
-2 tr_z (\delta_\e AF )
\w^{z^\prime}_3 ]\quad ,\label{d=6globalanomaly}
\ee
that  has to be augmented by the gauge-invariant terms
\beq
&&\Delta {\cal{A}}_\e =
A c_r^z c^{r z^\prime} tr_z [\delta_\e A_\m F_{\n\r} ]
tr_{z^\prime} [\bar{\l}^\prime \g^{\m\n\r} \l^\prime ] \nonumber\\
&&+ B c_r^z c^{r z^\prime}tr_z [\delta_\e A_\m \bar{\l}]
\g^{\m\n\r} tr_{z^\prime} [\l^\prime F^\prime_{\n\r} ]\nonumber\\
&&+ C c_r^z c^{r z^\prime } tr_z [\delta_\e A_\m \bar{\l}  ]
\g_\n  tr_{z^\prime} [\l^\prime F^{\prime \m\n } ]\quad ,
\label{completesusyanomaly}
\eeq
with coefficients satisfying the relations
\be
A+B =i \quad, \qquad C=4A - 2B \quad .\label{relations}
\ee
As pointed out in the previous Section, these leave one undetermined parameter, 
in agreement with the fact that the 
anomaly is defined up to the variation of a local
functional. Indeed, the variation
\be
\delta [(\bar{\l}\g^\a \l^\prime )(\bar{\l} \g_\a \l^\prime )]\label{lambda3}
\ee
yields the sum in eq. (\ref{completesusyanomaly}) with $A+B=0$ and
$C=4A-2B$, so that adding it to ${\cal A}_\e$ 
effectively does not affect eqs. (\ref{relations}). 
The main result, anticipated in \cite{wzsusy}, 
is that in six dimensions the last of eqs. 
(\ref{wesszumino}) generates terms containing one derivative and  
four gaugini, that cancel only using the Dirac equation
$\g^\m D_\m \l =0 $.  To reiterate, in six dimensions 
the Wess-Zumino consistency
conditions for an anomalous Yang-Mills theory close only on shell. 

The freedom of adding to the anomaly a $\delta \l^4$ term is another, related 
feature of the six-dimensional case that we would like to stress.
Indeed, one can easily observe that in $D$ dimensions
the effective weight\footnote{As usual in supersymmetric theories, this is 
1 for derivatives, 1/2 for fermi fields, and 0 for bose fields.}
of the supersymmetry anomaly is $(D-1)/2$, while 
that of the supersymmetry variation in (\ref{lambda3})
is $5/2$. Therefore, the undetermined $\l^4$ contribution is present only 
in six dimensions. Moreover, in all cases the breaking of 
supersymmetry affects the algebra on the gaugino,
that as usual closes only on the field equation.

All this has direct implications for six dimensional $(1,0)$ supergravity.
In this case, the Wess-Zumino conditions include additional contributions,
and become
\beq
\delta_\e {\cal{A}}_\L = \delta_\L {\cal{A}}_\e \quad ,\nonumber 
\eeq
\be
\delta_{\e_1} {\cal{A}}_{\e_2} -\delta_{\e_2} {\cal{A}}_{\e_1}=
{\cal A}_{\tilde{\e}}+{\cal{A}}_{\tilde{\L}} \quad .\label{newwz}
\ee
Again, they are only satisfied on-shell and, 
more precisely, if $\a=0$ one obtains
\beq
& & (\delta_{\e_1}{\cal{A}}_{\e_2} 
-\delta_{\e_2}{\cal{A}}_{\e_1})_{extra}
=\nonumber\\
& & c \cdot c^\prime tr \lbrace
\frac{1}{8}(\bar{\e}_1 \g_\sigma \e_2 )(\bar{\l}\g_\m \l^\prime )
(\bar{\l}\g^\m \g^\sigma
[eq.\l^\prime ]_{\a =0})\nonumber\\
& & -\frac{1}{16}
[\bar{\e}_1 \g_{\sigma\delta\tau} \e_2 ]_i \lbrace [\bar{\l}\g^\tau
\l^\prime ]_i (\bar{\l}\g^{\sigma\delta} [eq.\l^\prime ]_{\a =0})\nonumber\\
&&+[\bar{\l}\g^\tau \l ]_i (\bar{\l}^\prime \g^{\sigma\delta}
[ eq. \l^\prime ]_{\a =0}) \rbrace \rbrace \quad ,\label{openwz}
\eeq
where the terms within square brackets are fermionic bilinears with a
non-trivial
symplectic structure defined by the relation
\be
[\bar{\chi}\Psi ]_i = \sigma_{i a}{}^b \bar{\chi}_b \Psi^a \quad .
\ee

Still, the identity
\beq
&&  c \cdot c^\prime  tr \lbrace 
\frac{e}{8}(\bar{\e}_1 \g_\sigma \e_2 )( \bar{\l} \g_\m \l^\prime )
(\bar{\l}\g^\m \g^\sigma
[eq.\l^\prime ]_{\a})\nonumber\\
&& -\frac{e}{16}
[\bar{\e}_1 \g_{\sigma\delta\tau} \e_2 ]_i \lbrace [\bar{\l}\g^\tau
\l^\prime ]_i (\bar{\l}\g^{\sigma\delta} [eq.\l^\prime ]_{\a})\nonumber\\
&&+ [\bar{\l}\g^\tau \l ]_i (\bar{\l}^\prime \g^{\sigma\delta}
[ eq. \l^\prime ]_{\a }) \rbrace  \rbrace =\nonumber\\
&& c \cdot c^\prime  tr \lbrace 
\frac{e}{8}(\bar{\e}_1 \g_\sigma \e_2 )( \bar{\l} \g_\m \l^\prime )
(\bar{\l}\g^\m \g^\sigma
[eq.\l^\prime ]_{\a = 0}) \nonumber\\
& & -\frac{e}{16}
[\bar{\e}_1 \g_{\sigma\delta\tau} \e_2 ]_i \lbrace [\bar{\l}\g^\tau
\l^\prime ]_i (\bar{\l}\g^{\sigma\delta} [eq.\l^\prime ]_{\a = 0})\nonumber\\
&&+[\bar{\l}\g^\tau \l ]_i (\bar{\l}^\prime \g^{\sigma\delta}
[ eq. \l^\prime ]_{\a = 0 }) \rbrace \rbrace \nonumber\\
&& + \frac{e \a}{8}
\frac{c_r^z c^{r z^\prime}c_s^{z^\prime}c^{s z^{\prime\prime}}}{v_t c^{t z^\prime}}
tr_{z,z^\prime ,z^{\prime\prime} }
[\bar{\e_1}\g^{\m\n\r}\e_2 ]_i [\bar{\l}\g_\m \l ]_i \cdot \nonumber\\
&&[\bar{\l}^\prime \g_\n \l^\prime ]_j [\bar{\l}^{\prime\prime}\g_\r
\l^{\prime\prime} ]_j \label{magic}
\eeq
implies that the last term should somehow be generated in the anomaly, 
if the Wess-Zumino condition is to close for any value of $\a$. 
In the presence of ${\cal L}_{\l^4}$, however, the anomaly is modified by eq.
(\ref{extraanomaly}). Using the field equation of the 
gaugini for arbitrary $\a$,
the last of eqs. (\ref{wesszumino}) on this term yields, 
\be
[\delta_{\e_1},\delta_{\e_2}]{\cal{L}}_{\l^4}=\delta_{\tilde{\e}} 
{\cal L}_{\l^4} +\delta_\a {\cal L}_{\l^4}\quad ,\label{extrawz}
\ee
where $\delta_\a$ is the transformation defined in eq. (\ref{centralcharge}).
The end result is that this transformation exactly equals the last term in 
eq. (\ref{magic}), and
thus one can understand the rationale behind the occurrence of the 2-cocycle 
in the 
algebra on the gaugini: it lets the Wess-Zumino conditions close precisely on 
the field equations determined by the algebra.  Since 
the Wess-Zumino conditions need only the equation of the gaugini, 
only these fields sense the additional transformation.

To summarize, in six dimensions the Green-Schwarz mechanism exhibits some
relevant novelties, since the Green-Schwarz term is in 
this case $B \wedge F \wedge F$.
Although generated by a one loop contribution from 
the string viewpoint, in the low-energy supergravity this introduces
tree-level anomalies that are compensated by fermion loops. 
This fact has important consequences for the low-energy couplings, 
that play a central role in perturbative
six-dimensional type-I vacua and in recent 
discussions of string duality \cite{dmw},
some of which we have tried to elucidate.


\begin{thebibliography}{99}
\bibitem{gs} M.B. Green and J.H. Schwarz,	Phys. Lett. 149B (1984) 117.
\bibitem{r-d} S. Randjbar-Daemi, A. Salam, E. Sezgin and J. Strathdee,
Phys. Lett. 151B (1985) 351.
\bibitem{bs}{M. Bianchi and A. Sagnotti, Phys. Lett. B247
(1990) 517; Nucl. Phys. B361 (1991) 519;\\
E. Gimon and J. Polchinski, hep-th/9601038;\\
A. Dabholkar and J. Park, hep-th/9602030, 9604178;\\
E. Gimon and C.V. Johnson, hep-th/9604129;\\
M. Berkooz, R. Leigh, J. Polchinski, J.H. Schwarz, N. Seiberg and E. Witten,
Nucl. Phys. B475 (1996) 115;\\
J. Blum and A. Zaffaroni, Phys. Lett. B387 (1996) 71;\\
J. Blum, Nucl. Phys. B486 (1997) 34;\\
A. Sen, Nucl. Phys. B475 (1996) 562, Phys. Rev. D55 (1997) 7345, hep-th/9709159;\\ 
J. Blum and K. Intriligator, hep-th/9705030, 9705044;\\
J. Blum, K.R. Dienes, hep-th/9707148, 9707160;\\ 
K. Dasgupta and S. Mukhi, Phys. Lett. B385 (1996) 125;\\
C. Angelantonj, M. Bianchi, G. Pradisi, A. Sagnotti and Ya.S. Stanev,
Phys. Lett. B387 (1996) 743.\\
For a review, see: L.E. Iba\~nez and A.M. Uranga, hep-th/9707075.}
\bibitem{cargese} {A. Sagnotti, {\it in} 
``Non-Perturbative Quantum 
Field Theory'', eds. G. Mack et al (Pergamon Press, 1988), p. 521.}
\bibitem{as} {A. Sagnotti, Phys. Lett. 294B (1992) 196.}
\bibitem{frs}{S. Ferrara, F. Riccioni and A. Sagnotti, hep-th/9711059.}
\bibitem{romans} L.J. Romans,  Nucl. Phys. B276 (1986) 71.
\bibitem{ns1} H. Nishino and E. Sezgin, Nucl. Phys. B144 (1986) 353.
\bibitem{ns2} H. Nishino and E. Sezgin, hep-th/9703075.
\bibitem{fms} S. Ferrara, R. Minasian and A. Sagnotti, Nucl. Phys.
              B474 (1996) 323.
\bibitem{wz}{J. Wess and B. Zumino, Phys. Lett. B37 (1971) 95.}
\bibitem{schwarz} J.H. Schwarz, Nucl. Phys. B226 (1983) 269.
\bibitem{pst}{P. Pasti, D. Sorokin, M. Tonin, Phys. Rev. D55 (1995) 6292.}
\bibitem{wzsusy}{
G. Girardi, R. Grimm and R. Stora,  Phys. Lett. B156 (1985) 203;\\
L. Bonora, P. Pasti and M. Tonin, Phys. Lett. B156 (1985) 341;\\
E. Guadagnini, K. Konishi and M. Mintchev,  Phys. Lett. B157 (1985) 37;\\
N.K. Nielsen, Nucl. Phys. B244 (1984) 499;\\
H. Itoyama, V.P. Nair and H. Ren, Nucl. Phys. B262 (1985) 317;\\
E. Guadagnini and M. Mintchev,  Nucl. Phys. B269 (1986) 543;\\
S. Ferrara, A. Masiero, M. Porrati and R. Stora,
Nucl. Phys. B417 (1994) 238.}
\bibitem{dmw}{E. Witten, Nucl. Phys. B460 (1996) 541;\\
M.J. Duff, R. Minasian and E. Witten, Nucl. Phys. B465 (1996) 413;\\
N.~Seiberg and E.~Witten, Nucl. Phys. B471 (1996) 121;\\
M.J. Duff, H. L\"u and C. Pope, Phys. Lett. B378 (1996) 101.}
\end{thebibliography}
\end{document}